\documentclass[12pt,letterpaper]{article}
\pdfoutput=1
\usepackage{amsmath,amssymb,calc, amsthm}
\usepackage{graphicx}
\usepackage[nosort]{cite}
\usepackage{epsfig,psfrag}

\usepackage{ulem}
\usepackage{color}
\definecolor{burgundy}{rgb}{0.8,.2,.2}

\makeatletter
\renewcommand\section{\@startsection {section}{1}{\z@}%
                                   {-3.5ex \@plus -1ex \@minus -.2ex}%nn
                                   {2.3ex \@plus.2ex}%
                                   {\normalfont\large\bfseries}}
\renewcommand\subsection{\@startsection{subsection}{2}{\z@}%
                                     {-3.25ex\@plus -1ex \@minus -.2ex}%
                                     {1.5ex \@plus .2ex}%
                                     {\normalfont\bfseries}}
\makeatother

\theoremstyle{plain}

\theoremstyle{definition}

\let\non\nonumber

\let\s=\sigma

\def\one{^{(1)}}

\newcommand{\bea}{\begin{eqnarray}}
\newcommand{\eea}{\end{eqnarray}}
\newcommand{\be}{\begin{equation}}
\newcommand{\ee}{\end{equation}}
\newcommand{\bma}{\begin{pmatrix}}
\newcommand{\ema}{\end{pmatrix}}

\newcommand{\Z}{{\mathbb Z}}
\newcommand{\R}{{\mathbb R}}

\newcommand{\e}{\epsilon}

\newcommand{\C}[1]{$(\ref{#1})$}

\def\IZ{\relax\ifmmode\mathchoice
{\hbox{\cmss Z\kern-.4em Z}}{\hbox{\cmss Z\kern-.4em Z}}
{\lower.9pt\hbox{\cmsss Z\kern-.4em Z}} {\lower1.2pt\hbox{\cmsss
Z\kern-.4em Z}}\else{\cmss Z\kern-.4em Z}\fi}
\def\IR{\relax{\rm I\kern-.18em R}}

\def\one{{\hbox{ 1\kern-.8mm l}}}

\def\Tr{{\rm Tr\,}}

\newlength{\bredde}
\def\slash#1{\settowidth{\bredde}{$#1$}\ifmmode\,\raisebox{.15ex}{/}
\hspace*{-\bredde} #1\else$\,\raisebox{.15ex}{/}\hspace*{-\bredde}
#1$\fi}

\newsavebox{\zzzbar}
\sbox{\zzzbar}
  {\setlength{\unitlength}{0.9em}
  \begin{picture}(0.6,0.7)
  \thinlines
  \put(0,0){\line(1,0){0.6}}
  \put(0,0.75){\line(1,0){0.575}}
  \multiput(0,0)(0.0125,0.025){30}{\rule{0.3pt}{0.3pt}}
  \multiput(0.2,0)(0.0125,0.025){30}{\rule{0.3pt}{0.3pt}}
  \put(0,0.75){\line(0,-1){0.15}}
  \put(0.015,0.75){\line(0,-1){0.1}}
  \put(0.03,0.75){\line(0,-1){0.075}}
  \put(0.045,0.75){\line(0,-1){0.05}}
  \put(0.05,0.75){\line(0,-1){0.025}}
  \put(0.6,0){\line(0,1){0.15}}
  \put(0.585,0){\line(0,1){0.1}}
  \put(0.57,0){\line(0,1){0.075}}
  \put(0.555,0){\line(0,1){0.05}}
  \put(0.55,0){\line(0,1){0.025}}
  \end{picture}}

\newcommand{\ena}{\end{eqnarray}}
\newcommand{\beqa}{\begin{eqnarray}}
\newcommand{\eeqa}{\end{eqnarray}}

\newfont{\goth}{ygoth.tfm scaled 1200}                   % gothic font (usual)
\def\e{\epsilon}
\def\th{\theta}

\def\s{\sigma}

 \numberwithin{equation}{section}

\def\1{{(1)}}
\def\2{{(2)}}
\def\3{{(3)}}

\def\1{{\bf 1}}

\begin{document}
\begin{titlepage}

\begin{center}

%{June 13, 2012}
\today
\hfill         \phantom{xxx}  EFI-12-35

\vskip 2 cm {\Large \bf  A New String in Ten Dimensions?}
%\vskip 0.3 cm {\Large \bf K\"ahler Potentials}\non\\
\vskip 1.25 cm {\bf  Savdeep Sethi\footnote{sethi@uchicago.edu} }\non\\
%{\vskip 0.5cm  $^{a}$ {\it Dept. of Physics, University of Cincinnati,
%Cincinnati, OH 45221, USA}\non\\
\vskip 0.2 cm
%{\it $^{a}$Max Planck Institute for Gravitational Physics, Am M\"uhlenberg 1, D-14476 Golm, Germany}\non\\ \vskip 0.2cm
 {\it Enrico Fermi Institute, University of Chicago, Chicago, IL 60637, USA}\non\\ \vskip 0.2cm
%$^{c}${\it Department of Mathematics, Duke University, Durham, NC 27708, USA\non\\}
%Valckenierstraat 65, 1018 XE Amsterdam, The Netherlands} \non\\}

\end{center}
\vskip 2 cm

\begin{abstract}
\baselineskip=18pt

I suggest the possibility of a new string in ten dimensions. Evidence for this string is presented  both from orientifold physics and from K-theory, along with a mystery concerning the M-theory description. Motivated by this possibility, some novel aspects of decoupling limits in heterotic/type I theories are described; specifically, the decoupled theory on type I $D$-strings is argued to be three-dimensional rather than two-dimensional. These decoupled theories provide the matrix model definitions of the heterotic/type I strings. 

%============================================================================================================================

%I propose the existence of a possible new string theory in ten dimensions. At the perturbative level, the theory is identical to type I string theory. Non-perturbatively, the  spectrum appears to be different. The strong coupling limit is not described by a known string theory. Evidence for this string is presented  both from orientifold physics and from matrix theory. Some novel aspects of decoupling limits in heterotic/type I theories are described; specifically, the decoupled theory on type I $D$-strings is argued to be three-dimensional rather than two-dimensional. 

\end{abstract}

\end{titlepage}

%\tableofcontents

\section{A New String?}
\subsection{Introduction}
\label{intro}

\begin{figure}[ht]
\begin{center}
%\[
%\mbox{\begin{picture}(200,220)(80,40)
\includegraphics[scale=0.75]{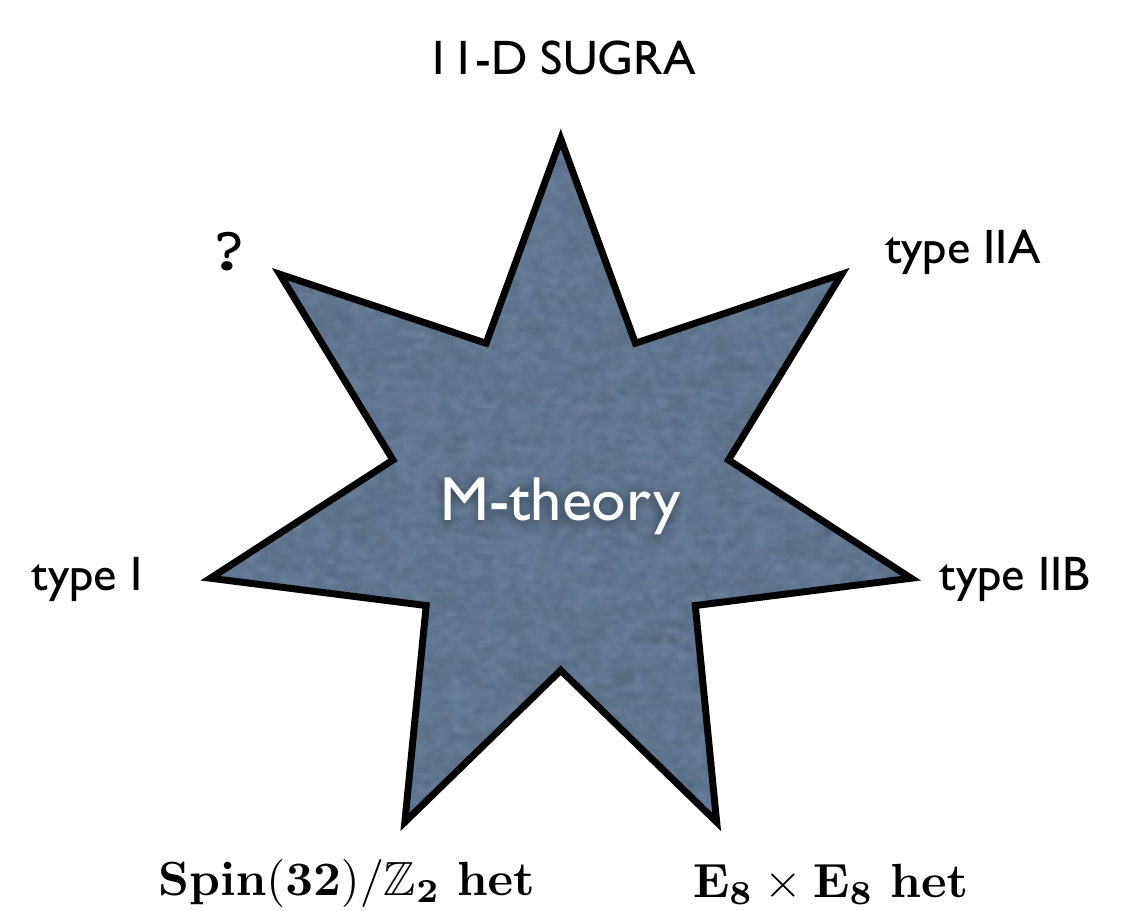}
%\end{picture}}
%\]
\vskip 0.1 in
 \caption{\it Perturbative limits of M-theory. } \label{perturblimits}
\end{center}
\end{figure}

While string theory is now a reasonably mature field, there are still surprises to be found even in fundamental string theory. There are five known perturbative strings in ten dimensions. The closed strings consist of the type IIA and type IIB theories which enjoy maximal supersymmetry~\cite{Cremmer:1978km, Schwarz:1983qr}. There are also $E_8\times E_8$ and $Spin(32)/\Z_2$ heterotic strings which preserve sixteen supersymmetries~\cite{Green:1984sg, Gross:1984dd}. Lastly, there is the type I open string which also preserves sixteen supersymmetries. These strings are all believed to be perturbative limits of M-theory. The final node of M-theory is its long wave-length expansion described by  eleven-dimensional supergravity.  This node has no coupling constant. This quite beautiful and unified picture of string theory is summarized in diagram~\ref{perturblimits}. 

One might wonder whether this diagram is complete. Could there be another spoke in this diagram? There are rather strong constraints on any such node. Its low-energy theory must be described by a supergravity theory. To have a scalar field that could serve as a string coupling constant, it must live in ten rather than eleven dimensions. All such low-energy supergravity theories with $16$ or $32$ supersymmetries are classified. Indeed one can go beyond the $2$ derivative long wave-length supergravity approximation and classify possible couplings to at least $8$ derivatives in theories with maximal supersymmetry~\cite{Green:1998by}. For theories with $16$ supersymmetries, much less is currently known about the higher momentum interactions beyond the $4$ derivative interactions, which are closely tied to anomaly cancelation. Indeed, anomaly cancelation is a very strong constraint on theories with $16$ supercharges. This constraint  determines the gauge group to be either $E_8\times E_8\ltimes \Z_2$, $Spin(32)/\Z_2$, $U(1)^{496}$ or $E_8\times U(1)^{248}$~\cite{Green:1984sg}. The latter two possibilities, $U(1)^{496}$ and $E_8\times U(1)^{248}$,  were ruled out in~\cite{Adams:2010zy}.

%\vskip 0.2in

This leaves very little room for anything new at weak coupling. Any new string will have to coincide at weak coupling with one of these known nodes unless there is a new mechanism for anomaly cancelation. Let us consider the type I string. The closed string sector of type I string theory can be viewed as an orientifold of type IIB string theory by world-sheet parity $\Omega$.    However, type IIB string theory has a RR axion $C_0$ which complexifies the type IIB string coupling $g_s=e^\phi$,
\be\label{iibcoupling}
\tau = C_0 + {i\over g_s}. 
\ee
The periodicity of $C_0$ is 
\be
C_0 \sim C_0 + 1,
\ee
with the convention that $\tau$ is identified under $SL(2,\Z)$:
\be\label{dualityactions}
\tau \, \rightarrow \, {a\tau + b\over c\tau +d}, \qquad (a,b,c,d)\in\Z, \quad ad-bc=1.
\ee
Conventional type I string theory corresponds to orientifolding at the point $C_0=0$. Under world-sheet parity $\Omega$, 
\be
C_0 \, \rightarrow \, -C_0, 
\ee
so there is a second possibility of taking $C_0={1\over 2}$ which is consistent with the orientifold action. At the perturbative level, there is no obvious way to distinguish this string from the type I string in flat space.  At low-energies, this theory should be described by type I supergravity in ten dimensions.

Clearly, the possibility of a new string is a highly speculative proposition. While the choice $C_0={1\over 2}$ seems to exist in perturbative string theory, there are several possibilities non-perturbatively: the theory might be inconsistent; the choice of $C_0$ might not be measurable and the theory might end up equivalent to conventional type I string theory; the non-perturbative definition might not fit into the web of M-theory backgrounds, like massive type IIA supergravity; or this might define a distinct consistent background. It is difficult to decide between these options without direct data about how $C_0$ affects the spectrum of $D$-branes and the structure of orientifold planes. Taking an optimistic attitude, we will explore some aspects of the theory which suggest significant differences from type I in the spectrum of $D$-branes. % and in the strong coupling limit. 
%that the   
 %A However, the theory appears to be different non-perturbatively. It differs 
% spectrum of $D$-branes and its strong coupling limit from conventional type I. 

Discrete choices like the value of $C_0$ which do not affect closed string perturbative physics, but do affect non-perturbative closed string physics (and possibly  open string physics) appear in a number of situations in field theory and string theory.  For example, Type I compactifications with gauge bundles that do not admit vector structure involve a discrete choice similar to the one proposed here~\cite{Bianchi:1991eu, Sen:1997pm, Witten:1997bs}. Those type I compactifications can be described as orientifolds of type IIB string theory with a ${1\over 2}$ unit of NS $B_2$-field threading a $2$-cycle. The simplest example is compactification on a torus $T^2$. Much like $C_0$, world-sheet parity inverts $B_2$ allowing for $2$ invariant choices under the orientifold action. In this case, the topology of the gauge bundle correlates with the choice of $B_2$-field.  

A similar example is found by considering type I compactified on $T^4$ with a configuration that corresponds to a non-trivial ``quadruple'' of holonomies~\cite{Boer:2002yq}. This is a zero energy non-trivial configuration of $Spin(32)/\Z_2$ Wilson lines on $T^4$, which cannot be deformed to a trivial configuration while maintaining zero energy. The existence of such a vacuum is quite surprising since there is no currently known invariant, analogous to a Chern-Simons invariant or a Pontryagin class, that distinguishes the non-trivial vacuum from a configuration deformable to the trivial vacuum.  Viewed as a type IIB orientifold, this choice of gauge bundle correlates with a ${1\over 2}$ unit of $C_4$ flux through $T^4$~\cite{Keurentjes:2001cp}. Precisely like $C_0$, the $C_4$ RR potential is inverted by world-sheet parity permitting two invariant choices. 

Similarly, there appear to be two possible orientifolds of type IIB on any six manifold, giving two type I compactifications. The two flavors of compactification  to four dimensions are distinguished by ${1\over 2}$ unit of NS $B_6$-potential threading the six manifold~\cite{Morrison:2001ct}. Alternately, one can view the two cases as differing by an expectation value of the  axion dual to the four-dimensional NS $B_2$-field. While this axion certainly affects non-perturbative amplitudes, whether the gauge-bundle can directly detect the expectation value remains mysterious. Other examples of discrete choices changing non-perturbative physics are found in strings propagating on gerbes~\cite{Pantev:2005rh, Pantev:2005wj}; for a review and further references on this topic, see~\cite{Sharpe:2010iv}.

%The outline for the remainder of this paper is as follows: in section~\ref{orientifold}, I will show that there are two possible orientifolds of type IIB string theory in ten dimensions in a more formal way. The argument, based on equivariant K-theory, bolsters the more straightforward physical argument given above. The effective action together with the spectrum of D-branes of the resulting theory are described.  In section~\ref{matrixdef}, I consider the Matrix Theory description of both type I string theory and this new string. This approach provides a non-perturbative definition of the light-cone quantized version of the string. It will also allow us to examine the strong coupling limit which is quite distinct from heterotic string theory.   

\subsection{The spectrum}

The most basic question that comes to mind is how to distinguish this new string from the conventional type I string. To answer this question, it is useful to recall the M-theory description of type I. We can obtain the M-theory description by compactifying type I on a circle and T-dualizing this circle. To obtain a perturbative compactification with no dilaton gradient~\cite{Polchinski:1996fk}, it is convenient to turn on a Wilson line on the circle breaking the space-time gauge group $G$ to a group with local structure 
\be\label{unbrokengauge}
G= SO(16) \times SO(16). 
\ee 
The resulting T-dual theory is type I$^\prime$, which is type IIA string theory on ${S}^1/\Omega\Z_2$. Let us use $\th_2$ as a coordinate for the $S^1$ factor. The quotient action inverts $\th_2$. There are two $O8^-$-planes localized at the two-fixed points of the $\Z_2$ action. By distributing an equal number of $D8$-branes at both ends of the interval, the charge cancelation is point-wise and the gauge group is~\C{unbrokengauge}. 

\begin{figure}[ht]
\begin{center}
\[
\mbox{\begin{picture}(200,100)(110,40)
\includegraphics[scale=1]{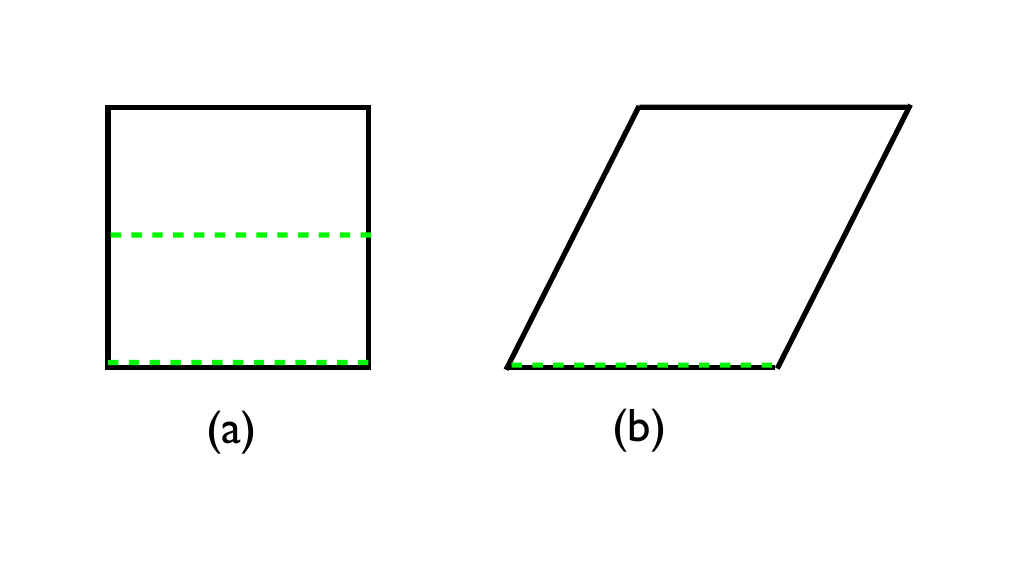}
\end{picture}}
\]
\vskip 0.1 in
 \caption{\it (a) Conventional type I$^\prime$ string theory with two boundaries. (b) The quotient~\C{definequotient}\  produces a M\"obius strip with one boundary.   } \label{tori}
\end{center}
\end{figure}

Taking the strong coupling limit gives M-theory compactified on the cylinder,  
\be
S^1 \times S^1/{\cal I},
\ee 
with coordinates $(\th_1, \th_2)$. The quotient ${\cal I}$ is a  $\Z_2$ orientifold action that inverts both the M-theory $3$-form potential $C_3$ as well as $\th_2$. This quotient produces two disjoint boundaries on which $SO(16)$ gauge bosons are localized. This geometry is depicted in diagram (a) of figure~\ref{tori}. Since the geometric quotient action is parity odd, the additional action on the M-theory $3$-form is required. This is the M-theory description of type I compactified on a circle.\footnote{Some interesting concerns about the existence of type I string theory beyond perturbation theory are described in~\cite{Copsey:2013jla}. We will work under the assumption that the theory exists non-perturbatively.} 

We can conveniently view this background as a quotient of M-theory on $T^2$, which describes type IIB string theory. This will be useful for comparing the spectra of type I against our proposed string. Let us start by recalling how the parameters and BPS spectra of M-theory on $T^2$ match the parameters and spectra of type IIB on a circle. We identify the complex structure of the torus with the type IIB coupling~\C{iibcoupling}. The area of the torus determines the size, $R_B$, of the IIB circle via the relation,
\be
 R_B = {\ell_p ^3\over A},
\ee 
where $\ell_p$ is the $11$-dimensional Planck scale. The decompactification limit corresponds to $A\rightarrow 0$. For the torus metric, we take
\bea\label{torus}
&&ds^2 = {A\over  \tau_2} dz d\bar z, \\
&& z= \th_1+i \th_2 \sim z +  (\hat{n}+\hat{m}\tau), \label{ident}
\eea
where $ A$ is the torus volume and $(\hat{n},\hat{m})$ are integers. 
%\be
%ds^2 = {A\over \tau_2} |d\th_1 + \tau d\th_2|^2, 
%\ee
%where $\th_1\sim \th_1+2\pi, \quad \th_2\sim \th_2+2\pi$. 

The Kaluza-Klein spectrum consists of point particle excitations in $9$-dimensions with masses, 
\be\label{masses}
M_{(n,m)}^2 = {1\over A \tau_2 } \left[ n^2\tau_2^2 + (m-\tau_1 n)^2 \right].
\ee
Each of these modes is BPS. The $(n,m)$ excited mode has charge $(n,m)$ with respect to the $9$-dimensional Kaluza-Klein gauge potentials 
\be\label{potentials}
  g_{\mu\th_1} -{\tau_1\over \tau_2} \, g_{\mu\th_2}, \qquad {1\over \tau_2} g_{\mu\th_2}.  
\ee
From the type IIB perspective, these modes arise from an $(n,m)$ string wrapping the IIB circle of size $R_B$. The $(1,0)$ string is the $D$-string while the $(0,1)$ string is the fundamental string. As $\tau_1$ interpolates from $0$ to $1$, a $(1,0)$ string turns into a $(1,-1)$ string.

The quotient action that produces type I acts on the M-theory torus with $\tau_1=0$ as follows:
\be\label{definequotient}
{\cal I}: (\th_1, \th_2, C_3) \rightarrow (\th_1, -\th_2 , -C_3). 
\ee
This projection kills the massless $g_{\mu \th_2}$ gauge potential. Invariance under the quotient action forces modes with momentum $(0,m)$ to appear in combination with modes of momentum $(0,-m)$. This is how M-theory reflects the absence of stable fundamental string winding modes in type I string theory, or equivalently, the absence of stable fundamental string momentum modes in the T-dual type I$^\prime$ theory. There are, however, stable BPS excitations coupling to $g_{\mu\th_1}$. In type I string theory, these are modes of the $D$-string wrapping the $R_B$ circle.  

How is the new string theory realized in M-theory? This is a difficult question. The ideal approach is to study the perturbative orientifold with $C_0={1\over 2}$, and deduce its $D$-brane spectrum and behavior under T-duality. Nevertheless, without that data we can try to proceed with some reasonable guesses about how the theory might behave. The T-duality that goes from type I to type I$^\prime$ is a basic closed string T-duality. It is unlikely to be affected by a RR potential like an expectation value for $C_0$. Therefore the right quotient action should include  $\th_2 \rightarrow -\th_2$ in the M-theory description. On the other hand, the nature of the orientifold planes produced by T-duality can differ from the two $O8^-$-planes that appear for the conventional type I case. There is already considerable evidence that this happens for the orientifold $T^4/\Omega$ in the presence of a discrete $C_4$ potential through the $T^4$~\cite{Keurentjes:2001cp}. It is quite critical to understand how this happens directly in perturbative string theory since the rank of the gauge group is reduced by $8$ in that example. If a similar rank reduction were happening in our case, it would suggest a different mechanism for anomaly cancelation.  

Let us proceed keeping in mind this caveat about the orientifold planes. Note that the torus identifications~\C{ident}\ are invariant under~\C{definequotient}\ for both $\tau_1=0$ and $\tau_1=1/2$. It is natural to guess that the latter case corresponds to the new string. This possibility is depicted in case (b) of figure~\ref{tori}. 

However, there is a puzzle. Instead of a cylinder, the quotient action~\C{definequotient}\ produces a space with one boundary; indeed, it produces a M\"obius strip. In the weak coupling limit, we expect two independent sets of $SO(16)$ gauge bosons localized at the ends of an interval. Yet in this case, we would seem to have one independent set. This suggests something unusual is going on: either the quotient action is more complicated than~\C{definequotient}, or perhaps the rank is reduced, and there is some alternate mechanism for anomaly cancelation. This would have to come from the RR potential modifying couplings on the orientifold planes. An alternate possibility is that the resulting type I$^\prime$ background might not be perturbative, much like the T-dual of type I with unbroken gauge symmetry. The boundary could then support a more exotic $9$-brane defect. This is the most significant mystery concerning the proposed new string. Note that another interpretation for M-theory on a finite size M\"obius strip, suggested in~\cite{Dabholkar:1996uq}, is in terms of the $9$-dimensional CHL string~\cite{Chaudhuri:1995kx}.  

%might be in conjunction with orientifold planes is modified corresponding to a  more exotic $9$-brane defect. In that case, the resulting type I$^\prime$ background might not be perturbative much like the T-dual background to type I with unbroken gauge symmetry. 

%Since the rank of the gauge group must be $16$ for anomaly cancelation, it hard to see how M-theory could produce the right gauge group from this geometry without a rather exotic $9$-brane defect. A more reasonable interpretation for M-theory on a finite size M\"obius strip, suggested in~\cite{Dabholkar:1996uq}, is in terms of the $9$-dimensional CHL string~\cite{Chaudhuri:1995kx}.  

%which has a maximal $E_8$ unbroken gauge group   

%\vskip 0.2in

% However, there is some freedom about how the quotient acts in the hidden eleventh dimension. Since we really would like two independent sets of gauge bosons, it seems reasonable that the quotient should generate two boundaries like the conventional type I case. Let us consider the action,
%\be\label{newquotient}
%{\cal I'}: (\th_1, \th_2, C_3) \rightarrow (\th_1- {\th_2\over \tau_2}, -\th_2 , -C_3). 
%\ee
%This motion respects the torus identifications~\C{ident}. However, the fixed point set now consists of two disjoint boundaries located at $(\th_1, 0)$ and $(\th_1, \tau_2/2)$.  This choice is depicted in case (c) of figure~\ref{tori}. 

Under the quotient ${\cal I}$ of~\C{definequotient}\ with $\tau_1=1/2$, the gravitational BPS spectrum changes quite significantly. Invariance under~\C{definequotient}\ forces states with Kaluza-Klein momenta $(n,m)$ to appear in the combinations, 
\be
|(n,m)\rangle \pm |(n, n-m)\rangle. 
\ee
The possible surviving BPS states therefore have Kaluza-Klein momenta $(2m,m)$. These modes couple to the space-time gauge potential,
 $g_{\mu\th_1}$,
which is invariant under the action~\C{definequotient}. The mass of these excitations, given by
\be\label{masses}
M_{(2m,m)}^2 = {4m^2 \tau_2 \over A },
\ee
is larger than conventional $D$-string wrapping modes in type I string theory. This contrast in the spectrum, which agrees with expectations for the CHL string, provides a rather sharp distinction between this new proposed string and type I string theory.    

%both gauge potentials are affected by the quotient. There are still modes charged under $g_{\mu\th_1}$, but they have an energy higher than we would expect from the BPS bound. In a sector with even string winding, it might be possible to find BPS bound states  but not in the sector with a singly wound string or an odd number of windings.  Addressing the existence of possible BPS states in even charge sectors is an interesting dynamical problem. 

A similar phenomenon occurs for configurations that are magnetically charged with respect to the potentials~\C{potentials}.  In M-theory, these are Kaluza-Klein monopole configurations with respect to the two $U(1)$ isometries of the torus. In type IIB string theory prior to any projection, these are $(p,q)$ $5$-branes. In conventional type I string theory, the $D5$-brane survives the projection as a stable BPS configuration. With $\tau_1=1/2$, only even charge bound states survive. %the tension of this brane is shifted. 

%we no longer expect a BPS unit charge $D5$-brane. The case of even charge configurations is again an interesting dynamical question that requires further analysis. 

% In type IIB string theory, turning on this flux can still give a BPS state because the BPS bound itself includes an expectation value for $C_0$, and because there exists a $B_2$ potential. In our case, world-sheet parity projects out $B_2$ leaving a configuration with energy above the BPS bound for a single $D$-string. 

While the spectrum appears quite different from type I, it is important to stress that the collection of stable excitations is the same. Indeed, while the global structure of the space-time gauge group might be quite subtle to determine, the local structure is fixed by anomaly cancelation to be isomorphic to $SO(32)$.\footnote{At least under the assumption of no new anomaly cancelation mechanism.} The existence of the non-trivial homotopy groups
\be
\pi_7(SO(32))=\Z, \qquad \pi_8(SO(32))=\Z_2, \qquad \pi_9(SO(32))=\Z_2,
\ee   
still implies the existence of a gauge string, a gauge particle and a gauge instanton configuration~\cite{Sen:1998tt, Sen:1998ki, Witten:1998cd}. %However, the physics of these excitations is now quite different. 

It is natural to ask what is going on from the world-volume perspective of $D$-strings and  $D5$-branes. Type IIB $(n,m)$ strings can be viewed as bound states of $m$ units of electric flux on the world volume of $n$ coincident $D$-strings~\cite{Witten:1995im}. The $C_0$ potential plays the role of a theta-angle in the two-dimensional gauge theory. Setting $C_0=1/2$ creates $1/2$ unit of electric flux on the $D$-string world-volume. Because of the Chern-Simons coupling on type II $D$-branes proportional to, 
\be
\int C_{RR} \wedge \Tr e^{2\pi\alpha' F}, 
\ee
where $F$ is the field strength for the world-volume gauge-field, it is interesting to speculate that turning on the background $C_0=1/2$ might correlate with turning on a discrete theta-angle on the orientifolded $D$-brane world-volume.

%It is interesting to speculate that the shift in tension is, perhaps, caused by some kind of discrete theta-angle on the $D$-string and $D5$-brane world-volumes.

It is worth noting that there is a possible discrete theta-angle for type I $D5$-branes. The world-volume for $N$ $D5$-branes supports an $Sp(N)$ gauge group in a notation where $N$ denotes the rank of the group. We note that 
\be\label{sixtheta}
\pi_5 (Sp(N)) = \Z_2 , \qquad N\geq1.
\ee
providing a discrete choice. Turning on the background $C_0=1/2$ might therefore correlate with turning on the discrete theta-angle permitted by~\C{sixtheta}.

% Because of the Chern-Simons coupling on type II $D$-branes proportional to, 
%\be
%\int C_{RR} \wedge \Tr e^{2\pi\alpha' F}, 
%\ee
%where $F$ is the field strength for the world-volume gauge-field, it is interesting to speculate that turning on the background $C_0=1/2$ might correlate with turning on the discrete theta-angle permitted by~\C{sixtheta}.

A similar suggestion for $D$-strings also looks plausible. 
%looks more mysterious. 
%also looks appealing.
The world-volume of $N$ $D$-strings supports an $O(N)$ gauge theory with $(0,8)$ supersymmetry. The choice of theta-angles is classified by
\be
\pi_1(O(N)) = \Z_2, \qquad N>2. 
\ee 
%For the case $N=1$, $\pi_1(O(2))=\Z$. 
The case $N=1$ is a discrete gauge theory, while $\pi_1(O(2))=\Z$. Since the BPS spectrum appears to consist of only even $N$ states, there is always a candidate theta-angle. 
%On the other hand, we have seen that there are no BPS single $D$-string states so perhaps this is not so surprising. 
%The mystery concerns the $N=1$ case and how the tension might increase to agree with the spectrum~\C{masses}. This is the case that realizes a single heterotic string in conventional type I string theory~\cite{Polchinski:1996fk}. A possible resolution to this mystery goes as follows: imagine wrapping this $D$-string on a circle and T-dualizing to type I$^\prime$. Since there are no Wilson line moduli for this wrapped string, the result is a $D0$-brane bound to the $O8$-plane walls.  
%On the other hand, the BPS gravitational spectrum seen in M-theory consists of mobile $D0$-branes, which are able to probe the $S^1/\Z_2$ interval. The single bound $D0$-brane is not visible in this spectrum since it involves strong interactions with the boundary degrees of freedom. Away from the orientifold planes, $N$ mobile $D0$-branes support a $U(N)$ gauge symmetry. This suggests we identify the Kaluza-Klein momentum $n$ with $2N$. 
To determine how a discrete RR potential like $C_0$ in type I or $C_1$ in type I$^\prime$ modifies the $D$-brane world volume theory will really require a direct study  in perturbative string theory.  
%It will be very interesting to study the world volume theory for $D$-branes in this new string both from perturbative string theory, and in M-theory by studying $M2$ and $M5$-branes along with Kaluza-Klein monopoles in the quotient geometry with $\tau_1=1/2$. 

\subsection{Confirmation from K-theory}

Both RR charges and RR fluxes are classified by K-theory in weakly-coupled string theory. 
For the type I string on ${\R}^{10}$, the choice of fluxes is believed to be classified by the group $KO^{-1}(\R^{10})$ in the absence of $B$-fields~\cite{Moore:2000fk}. The discrete choice of $C_0$ proposed in section~\ref{intro}\ should appear in such a classification. Alternately, viewed as an orientifold of type IIB, the group of fluxes can be computed using the formalism of~\cite{Distler:2009ri}, which applies to a very general class of orientifolds. Actually it is important that RR fields be viewed as elements of differential K-theory rather than topological K-theory in order to understand the relation with $C_0={1\over 2}$ as a fixed point under the orientifold action~\cite{Freed:2000tt}; for an exposition, see~\cite{freedexposition}.

To compute the K-theory group, we follow~\cite{Moore:2000fk}\ noting that $KO^{-1}(\R^{10})$ is equivalent to the reduced K-theory group $\widetilde{KO}(S^1\times \R^{10})$. We are interested in fluxes that preserve the translational symmetries of $\R^{10}$ so this group reduces to  $\widetilde{KO}(S^1)=\Z_2$. We identify these two choices with the two choices for $C_0$ described in section~\ref{intro}. The K-theory classification of fluxes provides further confirmation that there is indeed another possible string theory in ten dimensions.  

\subsection{The strong coupling limit}\label{strong}

The strong coupling limit of conventional type I string theory is the $Spin(32)/\Z_2$ heterotic string. We can understand this duality from M-theory as a consequence of the $SL(2,\Z)$ duality group of M-theory on $T^2$, given in~\C{dualityactions}. %, which is compatible with the quotient action~\C{definequotient}. 
For $\tau_1=0$, there is a $\Z_2$ subgroup of the full $SL(2,\Z)$ generated by, 
\be
\tau \rightarrow -{1\over \tau}, 
\ee
which exchanges $S^1$ and $S^1/\Z_2$. This action inverts the type I string coupling which is identified with the underlying type IIB coupling, 
\be
g_I \,\, \rightarrow \,\, g_{het} = {1\over g_I},
\ee  
and exchanges type I string theory with the $Spin(32)/\Z_2$ heterotic string. This latter exchange can been seen by first reducing M-theory on $S^1/\Z_2$ to get the $E_8\times E_8$ heterotic string~\cite{Horava:1995qa}. The subsequent reduction on $S^1$  with a choice of Wilson line breaking the gauge group to $SO(16)\times SO(16)$ followed by T-duality now gives $Spin(32)/\Z_2$ heterotic string theory rather than type I string theory.

For the case $\tau_1=1/2$, the situation is different.  We again identify the string coupling of the new string, $g_s$, with the underlying type IIB coupling $1\over \tau_2$. We need to examine whether any $SL(2,\Z)$ action on $T^2$ can map a small value of $\tau_2$ (corresponding to strong coupling) to a large value (corresponding to some weakly coupled theory). This is only possible for rational $\tau_1$ under an $SL(2,\Z)$ action~\C{dualityactions}\ with 
\be
c\tau_1 + d=0, \qquad \tau_2 \rightarrow {1\over c^2 \tau_2}. 
\ee
One such action is given by,
\be \label{change}
TST^2S = 
 \left( \begin{array}{cc}
1 & -1  \\
2 & -1   \end{array} \right),
\ee 
with $SL(2,\Z)$ generators: 
\be T = 
 \left( \begin{array}{cc}
1 & 1  \\
0 & 1   \end{array} \right), \qquad S = 
 \left( \begin{array}{cc}
0 & -1  \\
1 & 0   \end{array} \right).
\ee 
This transformation preserves $\tau_1=1/2$ and sends $\tau_2 \rightarrow {1\over 4 \tau_2}$. The factor of $4$ fits with 
the mass spectrum given in~\C{masses}\ suggesting a light closed string governs the strong coupling limit. 

The change of basis for the torus implemented by~\C{change}\  is not a symmetry of the theory because of the orbifold action. Rather, it is a mapping of one theory at strong coupling to a possibly distinct theory with large $\tau_2$. Whether that theory is really weakly coupled depends on whether $\tau_2$ still corresponds to some effective coupling constant. To describe the resulting model, it is convenient to shuffle the $\tau$-dependence of the periods back into the torus metric and work with coordinates $(\hat\theta_1, \hat\theta_2)$ with canonical periods,
\be
\hat\th_1 \sim \hat\th_1 + \hat{n}, \quad \hat\th_2 \sim \hat\th_2 + \hat{m}, \quad \hat{n}, \hat{m}\in \Z, \qquad {\widehat ds}^2 = {A\over \tau_2} |d\hat\th_1 + \tau d\hat\th_2|^2,
\ee 
rather than~\C{torus}. In hatted coordinates, the quotient action~\C{definequotient}\ sends: 
\be\label{hatquotient}
{\cal I}: (\hat\th_1, \hat\theta_2, C_3) \,\rightarrow\, (\hat\th_1+\hat\th_2, -\hat\theta_2, -C_3). \ee  
The $SL(2,\Z)$ action~\C{change}\ then describes the change of coordinates, 
\be\label{changeperiods}
(\hat\theta_1, \hat\theta_2) \,\rightarrow\, (\tilde\th_1, \tilde\th_2) = (-\hat\theta_1- \hat\theta_2, 2\hat\theta_1+ \hat\theta_2),
\ee
needed to implement the possible strong-weak coupling duality: 
\be
\tau \,\rightarrow\, {\tau-1 \over 2 \tau-1}.
\ee
In terms of the tilde coordinates~\C{changeperiods}, the quotient action~\C{hatquotient}\ becomes:
\be\label{dualquotient}
{\tilde I}: (\tilde\th_1, \tilde\th_2, C_3) \rightarrow (-\tilde\th_1- \tilde\th_2, \tilde\th_2 , -C_3). 
\ee
The fixed set under $\tilde I$ consists of the single boundary: 
\be
2\tilde\th_1 = \tilde\th_2.
\ee
%These two curves are drawn in figure~\ref{fixedset}. The real mystery is whether this limit can be formulated in terms of some set of weakly-coupled degrees of freedom with a coupling constant proportional to ${1\over \tau_2}$. The BPS closed string states visible in~\C{masses}\ remain heavy in this limit; unless there are light boundary states, a closed string description would seem unlikely. That leaves either a point particle theory or another open string theory, with the latter possibility favored. In fact, it is possible that this might be a self-dual open string theory, though the strong and weak coupling descriptions look superficially quite different. If a new perturbative definition can be formulated for this limit then there might well be another node on figure~\ref{perturblimits}. 

%\begin{figure}[ht]
%\begin{center}
%\[
%\mbox{\begin{picture}(00,120)(110,35)
%\includegraphics[scale=0.75]{fixedset.pdf}
%\end{picture}}
%\]
%\vskip 0.1 in
% \caption{\it The two fixed sets of the action~\C{dualquotient}. } \label{fixedset}
%\end{center}
%\end{figure}

\subsection{A DLCQ description}

Matrix theory provides a non-perturbative definition of discrete light-cone quantized string theories~\cite{Banks:1996vh}. Perhaps its most elegant success is in providing a definition of string theory in ten flat space-time dimensions; particularly for the case of the type IIA string~\cite{Motl:1997th, Dijkgraaf:1997vv}. Less well explored but perhaps just as elegant is the non-perturbative formulation of type IIB string theory in terms of the field theory on $M2$-branes wrapping $T^2$~\cite{Sethi:1997sw, Banks:1996my}.  It is critical that this theory is $2+1$-dimensional with the $SL(2,\Z)$ symmetry of the torus making manifest the $SL(2,\Z)$ of type IIB string theory.  We will see that matrix theory implements strong-weak coupling dualities for string theories with $16$ supersymmetries in a quite interesting way.

\subsubsection{The $SO(32)$ strings}

A matrix description of the $Spin(32)/\Z_2$ heterotic string was proposed in~\cite{Krogh:1998vb, Krogh:1998rw}. This should also describe the type I string at strong coupling. Let us review that result which starts by considering the $Spin(32)/\Z_2$ heterotic string compactified on a light-like circle, 
\be
X^- \sim X^- + R, 
\ee
with $N$ units of momentum $P^+ = {N\over R}$. The parameters of the theory are $g_s^{het}$ and  $\ell_s$. Following~\cite{Seiberg:1997ad}, we make the light-like circle slightly space-like 
\be
(X^+, X^-) \sim (X^+, X^-) + (-\e^2 R, R), 
\ee
where eventually we will take $\e\rightarrow 0$. A large boost $(X^+ = \e {x^+}, X^- = {x^- \over \e})$ relates this almost light-like compactification to a space-like compactification with $x^1\sim x^1 + \sqrt{2} \e R$ where $x^\pm = {1\over \sqrt{2}} \left(x^0\pm x^1 \right).$

With no light-like Wilson line, T-dualizing the $x^1$ circle gives back  the $Spin(32)/\Z_2$ heterotic string on a circle with radius scaling like ${1\over \e}$. The new string coupling also scales like ${1\over \e}$ so an S-duality is needed to take us to a weakly coupled type I theory with a string coupling scaling like $\e$ and a string-frame metric:
\be\label{closedbac}
ds^2 \sim {\e R \over \ell_s} \left\{ -2 dx^+ dx^- + \sum_{i=1}^8 (dx^i)^2\right\}.
\ee
This is exactly the same set of arguments that lead to the matrix string description of type IIA string theory. The momentum modes become type I $D$-strings wrapping $x^1$ in the background~\C{closedbac}. The $D$-strings support a $1+1$-dimensional field theory with world-volume coordinates $(\s,\tau)$. Aside from the $Spin(1,1)$ Lorentz group, the world-volume global symmetry group is $Spin(8)_R \times Spin(32)_f$. The action is completely determined by demanding $(0,8)$ supersymmetry and an unbroken $Spin(32)_f$ flavor symmetry. The gauge group is $O(N)$ and the matter multiplets consist of $8$ scalar fields transforming in the $({\bf 8}_V, {\bf 1})$ representation of the global symmetry group, and the symmetric $2$-tensor of $O(N)$.  Associated to these bosons by supersymmetry are right-moving fermions also in the symmetric $2$-tensor of $O(N)$. There are also left-moving gauginos in the adjoint of $O(N)$ partnered to the vector multiplet. The $1$-$9$ strings give left-moving fermions in the $({\bf 1}, {\bf 32})$ and the fundamental of $O(N)$.

Ignoring numerical factors, the initial heterotic parameters map to the gauge theory parameters as follows, 
\be
\s \sim \s+\ell_s, \qquad g_{YM} = {1\over g_s^{het} \ell_s}.  
\ee
The weakly-coupled heterotic string is described by a matrix string, very much along the lines of the type IIA matrix string. The type I string should emerge when $g_s^{I} = {1\over g_s^{het}}$ becomes small. The type I string scale, 
$$(\alpha')^I = g_s^{het} (\alpha')^{het},$$ 
is held fixed.  In terms of the gauge theory, this corresponds to energies $E  <<{1\over \ell_s}$ which implies a reduction to a quantum mechanical theory. 

Now let us take a different approach.\footnote{This is the approach to which citation $[11]$ of~\cite{Krogh:1998vb}\ refers.} Instead of starting with the $Spin(32)/\Z_2$ heterotic string, start with the type I string on a light-like circle with a Wilson line breaking the gauge symmetry to $SO(16)\times SO(16)$. The Wilson line is needed to maintain perturbative control after a T-duality. We will follow precisely the same set of manipulations that lead to the matrix description of the type IIB string. Boosting to a space-like circle requires a T-duality to type I$^\prime$ then a lift to M-theory. The result is a decoupled theory of $N$ membranes wrapped on $S^1\times S^1/\Z_2 $ with coordinates $(\th_1, \th_2)$. The type I string coupling and string scale $(g_s^I, \ell_s^I)$ map to the cylinder radii,  
\be
g_s^I = {R_1 \over R_2}, \qquad (\ell_s^I)^2  = R R_2. 
\ee
The weakly coupled heterotic string should emerge in the limit where $(\alpha')^{het} = R R_1$ is fixed and $R_2\rightarrow 0$, while the weakly coupled type I string appears with $R_2$ fixed and $R_1\rightarrow 0$. 

Now we meet a puzzle. The two S-dual procedures for finding a non-perturbative definition of the type I/heterotic string do not commute. The first theory, supported on type I $D$-strings, is superficially $1+1$-dimensional while the membrane theory is $2+1$-dimensional.  This is not an immediate contradiction because we are interested in a very special subset of states with energies scaling like ${1\over N}$; however, it looks intuitively wrong. The membrane theory appears to have more degrees of freedom, which are critical for the S-duality of the underlying type IIB string. In this limit where we expect the membrane theory to describe the heterotic string, we can Kaluza-Klein reduce along $S^1/\Z_2$ to get a matrix string theory. 

Aside from the difference in Wilson lines, this is essentially the theory on $N$ type I $D$-strings. Now we can ask about the additional massive Kaluza-Klein states with energies of order ${1\over R_2}$. These are not stable states in the membrane theory but they exist! From the original space-time perspective these states have light-cone energies,
\be
P^- \sim {R\over (\alpha')^I} = {R\over g_s^{het} (\alpha')^{het}},
\ee
that correspond to massive perturbative type I modes, or non-perturbative heterotic states. This suggests that something interesting is happening with the type I $D$-string theory, which does not happen with type IIB $D$-strings. We can speculate about what the difference might be. Type I $D$-strings differ from type IIB $D$-strings in one sharp way: type I $D$-strings can break. This is the S-dual version of the observation that there are open $Spin(32)/\Z_2$ heterotic strings~\cite{Polchinski:2006fk}. Therefore, we should sum over all open string sectors in constructing the decoupled theory on type I $D$-strings. This is a collection of unstable massive modes that supplement the usual degrees of freedom of the $D$-string gauge theory. 

There is a puzzle with this proposal. The energy of the end points found in~\cite{Polchinski:2006fk}\ is of order ${1\over (g_s^{het})^2}$, while we find modes with energies of order ${1\over g_s^{het}}$. Regardless of the precise identification of these modes, what seems to be clear is that the decoupled theory on type I $D$-strings is really $2+1$-dimensional with a collection of massive modes surviving the decoupling limit and promoting the $1+1$-dimensional theory to a theory of membranes with boundary.

We can now attempt a non-perturbative DLCQ definition of the new string with a light-like Wilson line breaking the gauge group to $SO(16)\times SO(16)$: consider $N$ membranes on a torus with $\tau_1=1/2$ quotiented by the action~\C{hatquotient}. These are impurity models with localized degrees of freedom similar to those described in~\cite{Sethi:1997zza, Kapustin:1998pb}; in this case, a collection of chiral fermions supported on the boundary. Unlike the conventional type I/heterotic theory, the matrix description might be central to understanding whether the theory actually exists. What is really in order is a systematic study of membranes with boundaries, extending the work of~\cite{Bagger:2007vi, Aharony:2008ug, Berman:2009xd, Berman:2009kj}; for a review of membrane physics, see~\cite{Bagger:2012jb}.      

\subsubsection{The $E_8\times E_8$ string}

One final comment is in order. Much past work has been devoted to the matrix model describing the $E_8\times E_8$ string. The approach is to start with M-theory on $S^1/\Z_2$ and follow the boost argument to arrive once again at the decoupled theory supported on type I $D$-strings; for a review and references, see~\cite{Taylor:2001vb}.  This approach only makes visible an $SO(16)\times SO(16)$ gauge symmetry because of the required T-duality from type I$^\prime$  to type I. 

 A more natural approach is to start with the $E_8\times E_8$ heterotic string compactified on a light-like circle with no Wilson line. The same boost argument followed by a T-duality leads back to the $E_8\times E_8$ heterotic string on a large circle with a string coupling growing like ${1\over \e}$. A subsequent lift to M-theory gives the theory of $N$ membranes compactified on $S^1 \times S^1/\Z_2$ with coordinates $(\th_2, \th_1)$. The heterotic string parameters $(g_s^{E_8}, \ell_s^{E_8})$ map to the membrane theory parameters as follows:
\be
g_s^{E_8}={R_1\over R_2} , \qquad (\ell_s^{E_8})^2 = R R_2.
\ee 
The collection of chiral fermions supported at each wall have boundary conditions appropriate for an $E_8$ current algebra. Once again these two S-dual procedures do not commute: the first gives the superficially $1+1$-dimensional field theory on $D$-strings, while the second again gives a membrane theory. The equivalence of these two procedures again supports the claim that the decoupled theory supported on $D$-strings is secretly $2+1$-dimensional. 

%The matrix description should commute with strong-weak coupling duality. 

%at a time before an  understanding of how to systematically derive such descriptions existed~\cite{Seiberg:1997ad}. 

%\section{Orientifold Actions}
%\label{orientifold}

%There is a more formal way of characterizing orientifolds of type IIB string theory in ten dimensions using 

%\section{Matrix Definition}
%\label{matrixdef}

%\vskip 0.5in
%\newpage
\subsection*{Acknowledgements}

I would like to thank Dan Freed, David Kutasov, Travis Maxfield, Greg Moore, Ashoke Sen and Edward Witten for helpful feedback. I am particularly grateful to Angel Uranga for considerable correspondence about a mistake in the first version of this manuscript. S.~S. is supported in part by
NSF Grant No.~PHY-0758029 and NSF Grant No.~0529954. 

\newpage
%\appendix

%\bibliographystyle{amsunsrt-ensp}
%\bibliography{myrefs}

\ifx\undefined\bysame
\newcommand{\bysame}{\leavevmode\hbox to3em{\hrulefill}\,}
\fi

\end{document}